\documentstyle[12pt,epsf]{article}
\newtheorem{Thm}{Theorem}[section]
\newtheorem{Defi}[Thm]{Definition}
\newtheorem{Lem}[Thm]{Lemma}

\newenvironment{proof}{\noindent {\bf proof:}\newline}{\QED}
\newcommand{\QED}{{\hspace*{\fill}{\vrule height 1.8ex 
width 1.8ex }\quad} 
    \vskip 0pt plus20pt}
\newcounter{saveeqn}

\newcommand{\newsection}[1]{\section{#1}\setcounter{equation}{0}}
\newcommand{\eq}[1]{(\ref{#1})}

\newcommand{\be}{\begin{equation}}
\newcommand{\ee}{\end{equation}}
\newcommand{\bea}{\begin{eqnarray}}
\newcommand{\eea}{\end{eqnarray}}
\newcommand{\beann}{\begin{eqnarray*}}
\newcommand{\eeann}{\end{eqnarray*}}

\newcommand{\Hs}{{\cal H}}

\newcommand{\up}{\uparrow}
\newcommand{\down}{\downarrow}

\newcommand{\hc}{\mbox{h.c.}}
\newcommand{\cc}[1]{c^{\mathstrut}_{\mathstrut #1}}
\newcommand{\ccd}[1]{c^{\mathstrut\dagger}_{\mathstrut #1}}
\newcommand{\dd}[1]{d^{\mathstrut}_{\mathstrut #1}}
\newcommand{\ddd}[1]{d^{\mathstrut\dagger}_{\mathstrut #1}}

\def\A1n{A_1\otimes\cdots\otimes A_n}
\def\Bar{\overline}
\def\phi{\varphi}            % redefinition!
\def\epsilon{\varepsilon}    % redefinition!

\def\A{{\cal A}}  \def\C{{\cal C}}

\catcode`@=11
\def\@biblabel#1{#1.}
\def\ifundefined#1{\expandafter\ifx\csname
                        \expandafter\eat\string#1\endcsname\relax}
\def\atdef#1{\expandafter\def\csname #1\endcsname}
\def\atedef#1{\expandafter\edef\csname #1\endcsname}
\def\atname#1{\csname #1\endcsname}
\def\ifempty#1{\ifx\@mp#1\@mp} 
\def\ifatundef#1#2#3{\expandafter\ifx\csname#1\endcsname\relax
                                  #2\else#3\fi}
\def\eat#1{}
\newdimen\refskip  \refskip=0pt %\parskip
\def\@utfirst #1,#2\@ver                                         
   {\author#1,\ifx#2\@ut\afteraut\else\@utsecond#2\@ver\fi}
\def\@utsecond #1,#2\@ver                                         
   {\ifx#2\@ut\andone\author#1,\afterauts\else ,\
      \author#1,\@utmore#2\@ver\fi}
\def\@utmore #1,#2\@ver                                         
   {\ifx#2\@ut\and\author#1,\afterauts\else ,\
      \author#1,\@utmore#2\@ver\fi}
\def\authors#1{\@utfirst#1,\@ut\@ver}                              
\let\more\relax  % executed after every subitem
\catcode`@=12
\def\Bref#1 "#2"#3{\authors{#1}:\ {\it #2}, #3\more} 
\def\Gref#1 "#2"#3{\authors{#1}\ifempty{#2}\else:``#2''\fi, #3\more}
\def\Grefnt#1 "#2"#3{\authors{#1}, #3\more} %no title version of \Gref
\def\Jref#1 "#2"#3{\relax \authors{#1}:``#2'', \Jn{#3}}  
\def\Jrefnt#1 "#2"#3{\relax \authors{#1}, \Jn{#3}} % nt vers. of \Jref
\def\inPr#1 "#2"#3\relax{in: \authors{\eds#1}:{\it #2}, #3} 
\newcommand{\Jn}[4]{{\it#1}\ {\bf#2} (#3), #4}
\def\author#1. #2,{#2, #1.}
\def\sameauthor#1{\leavevmode$\underline{\hbox to 25pt{}}$}  
\def\and{, and }   \def\andone{ and } 
\def\noinitial#1{\ignorespaces}
\let\afteraut\relax        
\let\afterauts\relax       
\def\etal{\def\afteraut{, et.al.}\let\afterauts\afteraut
           \let\and,}
\def\eds{\def\afteraut{(ed.)}\def\afterauts{(eds.)}}

\newcommand{\Rl}{{\bf R}}

\newcommand{\idty}{{\rm 1\hspace{-3.4pt} I}}
\newcommand{\Cir}{{\cal C}}
                                        
\begin{document}
\thispagestyle{empty}
{Archived as {\tt cond-mat/9604043 }\hspace{\fill} Preprint NMBN12-95}
%\special{!userdict begin /start-hook{gsave 200 30 translate
%65 rotate /Times-Roman findfont 216 scalefont setfont
%50 0 moveto 0.8 setgray (DRAFT) show grestore}def end}\vspace{30pt}
\begin{center}
{\LARGE\bf On the flux phase conjecture at half-filling: an
improved proof}
\\[20pt]
{\baselineskip=14pt 
Nicolas Macris\\ Institut de Physique Th\'eorique\\ Ecole
Polytechnique F\'ederale de Lausanne\\ CH 1015 Lausanne\\
Switzerland\\ E-mail: {\tt macris@eldp.epfl.ch}\\[15pt] Bruno
Nachtergaele\\ Department of Physics\\ Princeton University\\
Princeton, NJ 08544-0708, USA\\ E-mail: {\tt
bxn@math.princeton.edu}\\[15pt] (December 14, 1995)\\[15pt]
}

{\bf Abstract}\\[20pt]
\end{center}

We present a simplification of Lieb's proof of the flux phase
conjecture for interacting fermion systems --- such as the Hubbard
model ---, at half filling on a general class of graphs.  The main
ingredient is a procedure which transforms a class of fermionic
Hamiltonians into reflection positive form.  The method can also be
applied to other problems, which we briefly illustrate with two
examples concerning the $t-V$ model and an extended Falicov-Kimball
model.


\vfill

\hrule width2truein
\smallskip
{\baselineskip=12pt
\noindent
Copyright \copyright\ 1995 by the authors. Faithful reproduction of
this article by any means is permitted for non-commercial
purposes.\par }

\newpage
 
\newsection{Introduction}\label{sec:intro}

The main purpose of this note is to give a simplified version of
Lieb's proof of the flux phase conjecture in \cite{Lie}, which at the
same time allows for some straightforward generalizations.  Those
readers who are mainly interested in the basic argument, rather than
in learning about the more general description of it, are advised to
think about a finite regular square lattice on a cylinder while
reading this and the next section.  Once the argument is properly
understood the generalizations become straightforward.

The physical context where the first conjectures appeared \cite{Wie}
is reviewed in \cite{Fra}. For a history and a more general
formulation of the problem we refer to the first mathematical studies
on the subject by Lieb \cite{Lieb} and Lieb and Loss \cite{LL}.

Consider a system of spinless fermions (adding spin poses no problems)
on a finite set of sites $\Lambda$, at half-filling, and with a
Hamiltonian of the form
\be
H=\sum_{x,y\in\Lambda}t_{xy}\ccd{x}\cc{y} + H_{\hbox{\small int}}
\label{hams1}\ee
Here, $t_{xy}$ is a hermitian matrix. We will explain later what kind
of interactions $H_{\mbox{\tiny int}}$ are allowed (see Section
\ref{sec:proof}). For now, let us just say that the usual on-site
Hubbard interaction is among them (i.e. for spin $1/2$ fermions one
takes $H_{\hbox{\small int}} = U
\sum_{x\in\Lambda}(n_{x,\up}-1/2)(n_{x,\down}-1/2)$), and that only
gauge invariant interactions will be considered. 

Let $\Gamma$ be the graph with set of vertices $\Lambda$ and the set
of edges $\{<x,y>\mid t_{xy}\neq 0\}$.  A {\it circuit\/} in the graph
$\Gamma$ is a finite sequence $\gamma=(x_1,\ldots,x_k)$ of distinct
vertices, with the property that $<x_i,x_{i+1}>$, for
$i=1,\ldots,k-1$, and $<x_k,x_1>$, are all edges in the graph.  By
representing the circuit as an ordered sequence we have implicitly
given it one of the two possible orientations (for $k>2$).

The ground state energy of \eq{hams1} depends on the, in general
complex, parameters $t_{xy}$ only through their modulus $\vert
t_{xy}\vert$, and the flux variables $\Phi_\gamma$, for circuits
$\gamma$, which are defined as follows:
\be
\Phi_\gamma=\sum_{i=1}^{n}\phi_{x_i,x_{i+1}}, \quad \bmod 2\pi
\label{phigamma}\ee
where $\gamma=(x_1,\ldots,x_n)$, and $t_{xy}=
\exp (i\phi_{xy})\vert t_{xy}\vert$.
This follows from \cite[Lemma 2.1]{LL} where it was proved that there
is a unitary transformation relating any two Hamiltonians with phases
$\{\phi_{xy}\mid <xy>\in\Gamma\}$ that satisfy \eq{phigamma} with the
same fluxes $\Phi_\gamma$ for all $\gamma$. This unitary
transformation is of the form
\be
\ccd{x} \mapsto e^{i\theta_x} \ccd{x},\quad 
\cc{x} \mapsto e^{-i\theta_x} \cc{x}
\ee
and is called a gauge transformation. We will often write
$\{\phi_{xy}\}$ instead of $\{\phi_{xy}\mid <xy>\in\Gamma\}$.

The {\it flux phase problem\/} can now be formulated as follows: for
fixed values of the moduli $\vert t_{xy}\vert$, find the phases
$\phi_{xy}$ (or, equivalently, the fluxes $\Phi_\gamma$) for which the
ground state energy of the Hamiltonian \eq{hams1} attains its minimal
value. We cannot solve this problem in general. In fact, we do not
expect that there is a simple solution in general.  We are looking for
a solution in terms of a {\it basic set of circuits\/} $\Cir$ (e.g.,
the plaquettes of the square lattice). The set $\Cir$ should be not
too large and consist only of ``simple'' circuits, so that the
solution ($\equiv$ the values of the fluxes through the circuits in
$\Cir$) can be easily described. On the other hand $\Cir$ should
contain enough circuits so that their flux uniquely determines
$\Phi_\gamma$ for all circuits $\gamma$.

\begin{Defi}\label{def:basiccircuits}
A set $\Cir$ of circuits in a graph $\Gamma$ is called a basic set of
circuits if for any two configurations of phases $\{\phi_{xy}\}$ and
$\{\phi_{xy}^\prime\}$ that produce the same fluxes $\{\Phi_\gamma
\mid \gamma\in \Cir\}$, there exists a gauge transformation relating
$\{\phi_{xy}\}$ and $\{\phi_{xy}^\prime\}$,
i.e. $\phi_{xy}^\prime=\phi_{xy} +\theta_y -\theta_x$, for some real
$\theta_x, x\in \Lambda$.
\end{Defi}

Lieb and Loss showed in \cite[Lemma 2.1]{LL} that the set of all
circuits of a graph satisfies Definition
\ref{def:basiccircuits}. Often, it is more convenient
to work with a rather small subset of the set of all circuits.  For
examples of good choices of the set $\Cir$ we refer to Section
\ref{sec:examples}.

The class of models that we treat in this paper is described by the
following two assumptions on the graph $\Gamma$ together with the
configuration of $\vert t_{xy}\vert$ 's associated with the bonds:
\begin{description}
\item{\bf A1.} All circuits $\gamma=(x_1,\ldots,x_n)$ in $\Gamma$
are of even length, i.e. $n=2k$. This is equivalent to requiring that
the graph $\Gamma$ is bipartite, but we will not use explicitly a
decomposition into two sublattices.
\item{\bf A2.} There is a {\it basic set of circuits\/}  $\Cir$ such that
for each $\gamma\in\Cir$ there is an embedding of the graph in
$\Rl^D$, for some $D$, such that there is a $D-1$-dimensional
reflection hyperplane $P$ not containing any vertex of $\Lambda$, with
the following properties:
\begin{enumerate}
\item The whole graph $\Gamma$, together with the configuration of 
$\vert t_{xy}\vert$'s, is invariant under reflection through $P$.
\item All circuits $\gamma\in\Cir$ that are 
intersected by $P$ (i.e. not all vertices are in one of the two
halfspaces) are, up to orientation, invariant under reflection through
$P$.  In particular, $\gamma$ is invariant under reflection through
$P$.
\end{enumerate}
\end{description}

The embedding of the graph in $\Rl^D$ used to describe Assumption A2 is
not essential. We only introduce it in order to simplify the description.

Before we can state our main result we have to say what we mean by
``flux configuration'' and ``canonical flux configuration''.

\begin{Defi}\label{def:fluxconfiguration} {\bf (Flux configuration)}
Let $\{\Phi_\gamma\}$ be a set of fluxes (i.e. real numbers mod
$2\pi$) through all circuits of the graph. We say that
$\{\Phi_\gamma\}$ is a flux configuration, if there exist a set of
phases $\{\phi_{xy}\}$ such that (1.2) holds for all $\gamma$.
\end{Defi}

\begin{Defi}\label{def:canonicalflux} {\bf(Canonical flux configuration)}
Assume that $\Gamma$ satisfies the assumptions A1-2.  
A flux configuration $\{\Phi_\gamma\}$ is called canonical if
there is a set $\Cir$ of basic circuits satisfying A2 and such 
that for all $\gamma\in \Cir$, $\Phi_\gamma=0$ if $\gamma$ has length
$2\bmod 4$, and $\Phi_\gamma=\pi$ if $\gamma$ has length $0\bmod 4$.
\end{Defi}

Note that it is {\it not\/} true, in general, that in a canonical flux
configuration {\it all\/} circuits satisfy $\Phi_\gamma=0$ if $\gamma$
has length $2\bmod 4$, and $\Phi_\gamma=\pi$ if $\gamma$ has length
$0\bmod 4$.

The arguments in Section \ref{sec:proof} will show that, for graphs
that satisfy the assumptions A1-2, there always exists a
canonical flux configuration.

\begin{figure}[t]
\label{fig:1}
\begin{center}
\epsfysize=5truecm
\centerline{\epsfbox{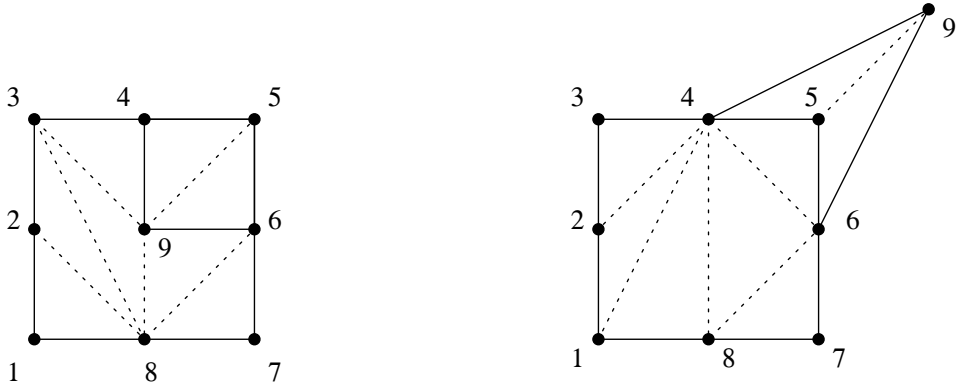}}
\parbox{14truecm}{\caption{\baselineskip=16 pt
Two different embeddings of a graph in the plane. The solid lines
indicate the edges. The dashed lines show a triangulation.
The Lieb-Loss flux through the circuit $(1,2,3,4,5,6,7,8)$ is $0$ 
for the first and $\pi$ for the second embedding.}}
\end{center}
\end{figure}

The definition of canonical flux configuration given here is different
from the one put forward in \cite{LL} for planar graphs embedded in
the plane.  A planar graph embedded in the plane can be triangulated
and Lieb and Loss \cite{LL} note that the number of triangles enclosed
by a circuit is independent of the triangulation, and they
define a flux configuration by putting a flux $\pi/2$ in each
triangle.  The resulting flux configuration for the original graph,
however, depends on the embedding in the plane one starts from (see
Figure 1 for an example).  Our definition is restricted to graphs that
have a basic set of circuits satisfying the assumptions A1 and
A2. They need not be planar, but, on the other hand many planar graphs
do not have a canonical flux configuration according to the definition
given here. Also we do not know whether there are graphs for which
different choices of $\Cir$ lead to different canonical flux
configurations.

Our main result is the following theorem.

\begin{Thm}\label{thm:main}
Under the assumptions A1 and A2 we have the following:\newline i)
There exists a configuration of phases $\{\phi_{xy}^{(c)}\}$ such that
the corresponding configuration of fluxes is a canonical
configuration.\newline 
ii) For the Hamiltonians \eq{hams1} we have
\be
\inf_{\{\phi_{xy}\}}\lambda_0(H(\{\phi_{xy}\}))=
\lambda_0(H(\{\phi_{xy}^{(c)}\}))
\ee
where $\lambda_0(H)$ denotes the smallest eigenvalue of $H$, i.e., 
canonical flux configurations minimize the ground state energy.
\end{Thm}

Quite generally we expect the energy minimizing flux configuration
to be unique up to gauge transformations, but
we have not studied the question of uniqueness. 
Non-uniqueness could arise in two ways. If there is
more than one canonical flux configuration the minimum will be
attained in both. The other possibility is that there is a
non-canonical minimizing flux configuration.

\newsection{Proof of the Main Result}\label{sec:proof}

\indent 
First, we only consider non-interacting spinless fermions. The
Hamiltonian is
\eq{hams1} with 
$H_{\mbox{\small int}}=0$. We will indicate at the end of this section
how spin and certain interactions can be included.

Statement i) of the Theorem \ref{thm:main} , for the case of planar
graphs, is a consequence of \cite[Lemma 2.2]{LL}.  For the more
general situation considered here i) will be a byproduct of the proof
of ii).

The main argument is an application of the Dyson-Lieb-Simon Lemma in
the following form.

\begin{Lem}\label{lem:DLS}
Let $A,B,C_1,\ldots,C_n$ be a collection of $d\times d$ complex
matrices ($n$ could be infinite) with the following properties: $A$
and $B$ are Hermitian, and for all $i$, $C_i$ is real and $\sum_i
C_i\otimes C_i$ is symmetric (as a $d^2\times d^2$ matrix). Let
$\lambda_0(A,B)$ denote the lowest eigenvalue of the matrix
\be
T(A,B)\equiv A\otimes\idty+\idty\otimes B \quad -\sum_i C_i\otimes C_i
\label{TAB}\ee
Then
\be
\lambda_0(A,B)\geq \frac{1}{2}\left( \lambda_0(A,\overline{A}
)+\lambda_0(\overline{B},B)\right)
\label{TABineq}\ee
where $\overline{A}$ denotes the matrix obtained from $A$ by complex
conjugation of the matrix elements. In particular
\be
\lambda_0(A,B)\geq \min\left(
\lambda_0(A,\overline{A}),\lambda_0(\overline{B},B)\right)
\ee
\end{Lem}

In the formulation of this lemma in \cite{DLS} the matrices $A$ and
$B$ are required to have real matrix elements. It is crucial for our
application that we consider complex matrices $A$ and $B$. This is a
straightforward extension. For a proof of Lemma \ref{lem:DLS} in the 
zero-temperature form stated here see \cite{LN}.

Before we can apply this lemma we have to bring the Hamiltonian into
the form \eq{TAB}. This will be achieved in three steps each
consisting of an elementary transformation.

Given a circuit of the set $\C$ we consider an embedding of the graph
in ${\bf R}^D$ and a reflection plane $P$ of the circuit (it exists by
assumption). This defines a left part $(L)$, a right part $(R)$, and a
set $(M)$ of vertices which belong to edges $<x,y>$ with $x\in L$ and
$y\in R$ or $x\in R$ and $y\in L$. The three steps are:
\begin{description}
\item{i)} A {\it Jordan-Wigner\/} type transformation,
\item{ii)} A {\it particle-hole\/} transformation,
\item{iii)} A {\it gauge\/} transformation.
\end{description}
We know from experience that one easily gets confused while performing
this sequence of transformations. Therefore, we now spell them out in
detail and indicate the purpose of each of them.

\noindent
{\it Step i):}
We introduce new operators $d^\#_x$ defined by
\be
\dd{x} = (-1)^{N_L}\cc{x}\ ,\quad
\ddd{x} = \ccd{x}(-1)^{N_L}\label{JW}
\ee
for all $x\in\Lambda$, and where $N_L$ is the total particle number in
the left half of the lattice, i.e., $N_L=\sum_{x\in L} \ccd{x}
\cc{x}$.
If one considers fermions with spin, $N_L$ has to be the total
particle number on the left, i.e., $N_L=\sum_{x\in L,\sigma} 
c^{\dag}_{x\sigma} c_{x\sigma}$. 
In one dimension, the transformation defined in
\eq{JW} is similar to the usual Jordan-Wigner transformation.
Strictly speaking however, even in one dimension, it is different. 
A slightly different transformations was employed previously by 
several authors, e.g., in \cite{FILS2}.  Note however that in \cite{FILS2} 
the paragraph about fermions contains a mistake. With the transformation
employed there the hopping terms on the right acquire the {\it opposite\/}
sign of the hopping terms on the left, and thus the Hamiltonian is 
{\it not\/} in reflection positive form.

Using the canonical anticommutation relations of the $c$ operators,
one easily finds that the $d$ operators satisfy the following algebra:
\beann
\left.\begin{array}{rcl}
\{\ddd{x}, \dd{y}\} &=& \delta_{xy}\\
\{\dd{x},\dd{y}\}=\{\ddd{x},\ddd{y}\} &=& 0
\end{array}\right\}
&&\mbox{if } x,y\in L \mbox{ or } x,y\in R\\
\left.\begin{array}{rcl}
[\ddd{x}, \dd{y}] &=& 0\\ {[\dd{x},\dd{y}]=[\ddd{x},\ddd{y}]}&=& 0
\end{array}\right\}
&&\mbox{if } x\in L, y\in R \mbox{ or } x\in R,y\in L
\eeann
The operators $d^\#_x$ acting on Fock space (associated to $\Lambda$)
 can be identified with operators of the form
\beann
d^\#_x\otimes\idty&& \mbox{ for } x\in L\\
\idty\otimes d^\#_x&& \mbox{ for } x\in R
\eeann
acting on the tensor product space $\Hs_L\otimes\Hs_R$, where each
factor corresponds to the Fock space associated to the left and right
parts of the lattice.  In terms of the $d^\#_x$ the Hamiltonian can be
considered as acting on $\Hs_L\otimes\Hs_R$ and takes the form
\bea
H&=&\sum_{x,y\in\Lambda}t_{xy}\ddd{x} \dd{y}\\ &=& \sum_{x,y\in
L}t_{xy}\ddd{x} \dd{y} + \sum_{x,y\in R}t_{xy}\ddd{x} \dd{y}
+\sum_{x,y\in M}t_{xy}\ddd{x} \dd{y}
\label{hamd}\eea
The third term of \eq{hamd} describes the interaction between the left
and the right half of the lattice and is of the tensor product form as
in \eq{TAB}.

\noindent
{\it Step ii):} The second step is a simple particle-hole
transformation on the right half of the lattice. i.e., for all $x\in
R$
\be
\dd{x}\mapsto \ddd{x},\quad \ddd{x}\mapsto \dd{x}
\ee
while the $d^\#_x$ with $x\in L$ remain unchanged.  The Hamiltonian
becomes
\bea
H&=& \sum_{x,y\in L}t_{xy}\ddd{x} \dd{y} + \sum_{x,y\in
R}(-\overline{t_{xy}})\ddd{x} \dd{y} \\ && +\sum_{x\in L, y\in
R}t_{xy}\ddd{x} \ddd{y} + \sum_{x\in R, y\in L}t_{xy}\dd{x} \dd{y}
\label{hamph}\eea

\noindent
{\it Step iii):} Finally we perform a gauge transformation with the
purpose of making the hopping matrix elements across the reflection
plane all negative. A transformation that achieves this is the
following:
\be
\begin{array}{rcl}
\ddd{y} &\mapsto& -e^{-i\phi_{xy}} \ddd{y}\\
\dd{y} &\mapsto& -e^{i\phi_{xy}} \dd{y}
\end{array}
\ee
for sites $y\in R$ which are connected to a site $x\in L$ (i.e. given
$y\in R$ there exist an $x\in L$ such that $t_{xy}\neq 0$).

\newcommand{\tp}{t^{\prime}}
\newcommand{\tpp}{t^{\prime\prime}}

Therefore we have a new set of hopping matrix elements $\{\tp_{xy}\}$
with {\it the same fluxes} as the original configuration
$\{t_{xy}\}$, (because a gauge transformation does not change the
fluxes) and $\vert \tp_{xy}\vert =\vert t_{xy}\vert$, in terms of
which the Hamiltonian is
\bea
H&=& \sum_{x,y\in L}\tp_{xy}\ddd{x} \dd{y} + \sum_{x,y\in
R}(-\overline{\tp_{xy}}) \ddd{x} \dd{y}\\ && -\sum_{x\in L, y\in
R}\vert\tp_{xy}\vert \ddd{x} \ddd{y} - \sum_{x\in R, y\in
L}\vert\tp_{xy}\vert
\dd{x} \dd{y} 
\label{finalham}\eea

We denote by $\Phi_L$ (respectively $\Phi_R$) the set of fluxes through 
basic circuits which are entirely in $L$ (respectively $R$), and by 
$\Phi_M$ the
flux configuration for basic circuits which have the same reflection
plane $P$. These fluxes refer to a particular orientation of the basic
circuits: first orient in an arbitrary way all circuits on the left,
and for the circuits on the right take the orientation opposite to the
one obtained by reflection of the left part. For the ones in the
middle choose an arbitrary orientation.  For a fixed configuration of
$\vert t_{xy}\vert$, the ground state energy depends on the phases
$\phi_{xy}$ only through the fluxes and we will denote this energy by
$E_0(\Phi_L,\Phi_M,\Phi_R)$.

We adopt the convention that the same set of fluxes $\Phi_L$ when it
appears as the third argument of $E_0$, assigns the flux to a circuit
on the right that is associated by reflection to the circuit on the
left. We can now state the basic lemma.

\begin{Lem}\label{lem:reflection}
Assume that the configuration $\{\vert t_{xy}\vert\}$ is invariant
under reflections. Then
\be
E_0(\Phi_L,\Phi_M,\Phi_R)\geq
\frac{1}{2}\left(E_0(-\Phi_R,\Phi_M^{(c)},\Phi_R) +
E_0(\Phi_L,\Phi_M^{(c)},-\Phi_L)\right)
\label{reflection}
\ee
where $\Phi_M^{(c)}$ is the canonical flux configuration through the
basic circuits intersecting $P$.
\end{Lem}
\begin{proof}
The proof is a direct application of Lemma \ref{lem:DLS} to the
Hamiltonian in the form \eq{finalham}, while carefully keeping track
of the flux configurations.  The operator $T(A,B)$ of \eq{TAB} is
given by
\beann
A&=&\sum_{x,y\in L}\tp_{xy}\ddd{x} \dd{y}\\ B&=&\sum_{x,y\in
R}(-\overline{\tp_{xy}})\ddd{x} \dd{y}\\
\sum_i C_i\otimes C_i&=& \sum_{x\in L, y\in R}
\vert\tp_{xy}\vert \ddd{x} \ddd{y}
+ \sum_{x\in R, y\in L}\vert\tp_{xy}\vert\dd{x} \dd{y}
\eeann
and $\lambda_0(A,\overline{A})$ is the ground state energy of the
Hamiltonian
\beann
T(A,\overline{A})&=&
\sum_{x,y\in L}\tp_{xy}\ddd{x} \dd{y}
+ \sum_{x,y\in L}\overline{\tp_{xy}}d^{\dag}_{r(x)} d_{r(y)}%\\ &&
-\sum_{x\in L, y\in R}\vert\tp_{xy}\vert \ddd{x} \ddd{y} - \sum_{x\in
R, y\in L}\vert\tp_{xy}\vert\dd{x} \dd{y}\\ 
&=& \sum_{x,y\in L}\tp_{xy}\ddd{x} \dd{y}
+ \sum_{x,y\in L}\tp_{xy} d^{\dag}_{r(y)} d_{r(x)}%\\ && 
-\sum_{x\in L, y\in R}\vert\tp_{xy}\vert \ddd{x} \ddd{y} - 
\sum_{x\in R, y\in
L}\vert\tp_{xy}\vert\dd{x} \dd{y}\\
\eeann
where $r(x)$ denotes the reflection of the site $x$ through $P$.  This
can be written back in the form of \eq{finalham} with a new
configuration of hopping matrix elements $\tpp_{xy}$ which do {\it
not\/}, in general, have the same fluxes as the original hoppings.  We
now determine the new configuration of fluxes
$(\Phi_L^{\prime\prime},\Phi_M^{\prime\prime},\Phi_R^{\prime\prime})$.

\begin{figure}[t]
\label{fig:2}
\begin{center}
\epsfxsize=14truecm
\centerline{\epsfbox{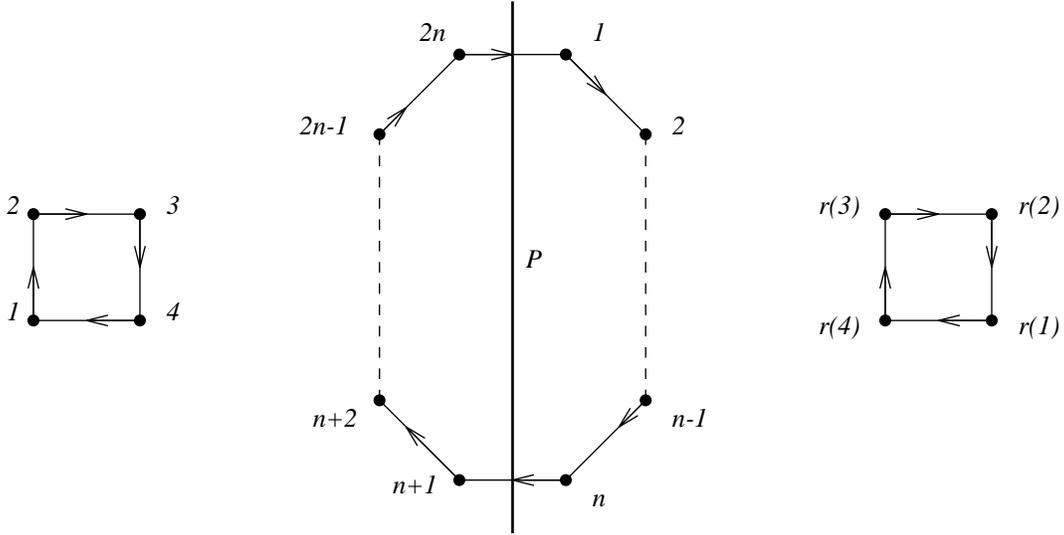}}
\parbox{14truecm}{\caption{\baselineskip=16pt 
By assumption, when a circuit is intersected by the 
reflection plane,
it is reflected into itself. Such circuits constitute the ``middle part''
of the graph. The figure also shows a circuit $(1,2,3,4)$ in the left part
of the graph and its reflection $(r(1),r(2),r(3),r(4))$ on the right.}}
\end{center}
\end{figure}

First, take a circuit in the middle part $\gamma=(x_1,...,x_{2n})$.
We label the vertices such that $x_1,\ldots,x_n\in R$, and
$x_{n+1},\dots,x_{2n}\in L$ (see Figure 2).  The edges intersected by
$P$ are $<x_{2n},x_1>$ and $<x_n, x_{n+1}>$.  The corresponding term
in the transformed Hamiltonian is
\bea
&&\sum_{i=n+1}^{2n-1} t_{x_i,x_{i+1}}^\prime d_{x_i}^\dagger
d_{x_{i+1}} +\sum_{i=1}^{n-1} (-\Bar{t_{x_i,x_{i+1}}^\prime})
d_{x_i}^\dagger d_{x_{i+1}}\\ &\quad& -\vert
t_{x_{2n},x_1}^\prime\vert d_{x_{2n}}^\dagger d_{x_1}^\dagger -\vert
t_{x_{n+1},x_n}^\prime\vert d_{x_{n+1}}^\dagger d_{x_n}^\dagger + \hc
\eea
and the corresponding flux is the original one
$\Phi_\gamma=\sum_{i=1}^{2n}
\phi_{x_i,x_{i+1}}=\sum_{i=n+1}^{2n-1}\phi_{x_i,x_{i+1}}^\prime
+\sum_{i=1}^{n-1}
\phi_{x_i,x_{i+1}}^\prime\bmod 2\pi$.
After reflection it becomes
\bea
&&\sum_{i=n+1}^{2n-1} t_{x_i,x_{i+1}}^\prime d_{x_i}^\dagger
d_{x_{i+1}} +\sum_{i=n+1}^{2n-1} t_{x_i,x_{i+1}}^\prime
d_{r(x_{i+1})}^\dagger d_{r(x_i)}\\ &\quad& -\vert
t_{x_{2n},x_1}^\prime\vert d_{x_{2n}}^\dagger d_{x_1}^\dagger -\vert
t_{x_{n+1},x_n}^\prime\vert d_{x_{n+1}}^\dagger d_{x_n}^\dagger +\hc
\eea
The new flux on $\gamma$ is $\Phi^{\prime\prime}_\gamma=
\sum_{i=n+1}^{2n-1}\phi_{x_i,x_{i+1}}^\prime+\sum_{i=n+1}^{2n-1}(\pi 
-\phi_{x_i,x_{i+1}}^\prime)\bmod 2\pi$, which is equal to $(n-1)\pi$,
i.e., the canonical flux through the circuit $\gamma$.

Next, we consider a circuit on the left $\gamma = (x_1,...,x_{2n})$,
oriented from $x_1$ to $x_{2n}$ ($n$ is an integer), and its
reflection $r(\gamma)=(r(x_1),...,r(x_{2n}))$ on the right oriented
from $r(x_{2n})$ to $r(x_1)$ (See Figure 2).  After the
transformations i)-iii) the corresponding terms in the Hamiltonian are
(with the convention $x_{2n+1}=x_1$)
\bea
A&=&\sum_{i=1}^{2n} t_{x_i,x_{i+1}}^\prime d_{x_i}^\dagger d_{x_{i+1}}
+ \hc
\label{3.2}\\
B&=&\sum_{i=1}^{2n} (-\Bar{t_{r(x_{i+1}),r(x_i)}^\prime})
d_{r(x_{i+1})}^\dagger d_{r(x_i)} + \hc
\eea
and the fluxes through $\gamma$ and $r(\gamma )$ are respectively
$\Phi_\gamma=\sum_{i=1}^{2n}\phi_{x_i,x_{i+1}}=
\sum_{i=1}^{2n}\phi_{x_i,x_{i+1}}^\prime$ and 
$\Phi_{r(\gamma)}=\sum_{i=1}^{2n}\phi_{r(x_{i+1}),r(x_{i})}=
\sum_{i=1}^{2n}\phi_{r(x_{i+1}),r(x_{i})}^\prime$. When we apply
Lemma 2.1 we have to replace $B$ by
\be
\bar A=\sum_{i=1}^{2n} \Bar{t_{x_i,x_{i+1}}^\prime}
d_{r(x_{i})}^\dagger d_{r(x_{i+1})} + \hc
=\sum_{i=1}^{2n}t_{x_i,x_{i+1}}^\prime d_{r(x_{(i+1)})}^\dagger
d_{r(x_{i})} + \hc
\ee
The new corresponding flux through $r(\gamma)$ is
$\Phi^{\prime\prime}_{r(\gamma)}=\sum_{i=1}^{2n}
(\pi-\phi_{x_i,x_{i+1}}^\prime)$, which is equal to $-\Phi_\gamma$. In
particular if $\gamma$ on the left has the flux $0$ or $\pi$ then
$r(\gamma)$ on the right has the same flux.

One argues similarly for $\lambda_0(\overline{B},B)$. This ends the
proof of Lemma \ref{lem:reflection}.
\end{proof}

\noindent
{\bf Proof of Theorem \ref{thm:main}} By assumption the configuration
$\{\vert t_{xy}\vert\}$ is invariant under reflections through all
reflection planes of the circuits in $\cal C$.  The crucial property
is that for each basic circuit there is a reflection plane that
intersects it and for which the conditions of Lemma
\ref{lem:reflection} are satisfied.
The theorem is then proved as an application of Lemma
\ref{lem:reflection}. The Lemma yields the existence of a
configuration of fluxes for which the ground state energy is at least
as low, while at the same time the new flux configuration is produced
from the old one by either
\be
(\Phi_L, \Phi_M, \Phi_R)\mapsto (\Phi_L,\Phi_M^{(c)}, -\Phi_L)
,\mbox{\ or\ } (\Phi_L, \Phi_M, \Phi_R)\mapsto (-\Phi_R,\Phi_M^{(c)},
\Phi_R)
\label{newLR}
\ee
In both cases the flux in all circuits intersected by $P$ becomes
canonical.  By the same argument as in \cite[Proof of Theorem
4.2]{DLS} or \cite{FILS}, one can now prove that the minimum is
attained in a canonical configuration by showing that, in an energy
minimizing configuration, the maximum number of circuits in $\Cir$
with canonical flux must be the total number of circuits in 
$\vert\Cir\vert$.
Let $\{\Phi_\gamma\}$ be a minimizing configuration with a given
number $N_c(\{\Phi_\gamma\})$ of circuits (in $\cal C$) with canonical
flux, and let $\gamma_0\in\Cir$ be a circuit that does not have
canonical flux in that configuration. Let $P$ be a reflection plane
leaving $\gamma_0$ invariant. After reflection the new configurations
in \eq{newLR}
%and \eq{newR} 
both have the same minimal energy.  Then writing
$\{\Phi_\gamma\}=(\Phi_L,\Phi_M,\Phi_R)$ $$ N_c(\Phi_L,\Phi_M^{(c)},
-\Phi_L) + N_c(-\Phi_R,\Phi_M^{(c)}, \Phi_R) = 2\left( N_c(\Phi_L,
\Phi_M, \Phi_R) + N_c(\Phi_M^{(c)}) - N_c(\Phi_M)\right) $$ As
$\gamma_0$ is a circuit in $M$ that does not have canonical flux in
$\Phi_M$ while in $\Phi_M^{(c)}$ it does (just like any other circuit
of $\Cir$ intersected by $P$) it is clear that $N_c(\Phi_M^{(c)}) -
N_c(\Phi_M) \geq 1$. We conclude that at least one of the new
minimizing configurations has strictly more circuits with canonical
flux than $\{\Phi_\gamma\}$.  This argument is then repeated until all
$\gamma\in\cal C$ have canonical flux.
\endproof

\noindent {\bf Remarks:}

\noindent {\it a) Finite temperatures.} Lemma 2.1 holds with $\lambda_0$
replaced by $-\log {\rm tr}\exp(-\beta H)$. Thus Lemma
 \ref{lem:reflection} holds also with the ground state energy $E_0$
 replaced by the free energy (at half filling), and of course its
 proof and the proof of theorem
\ref{thm:main} is the same.

\noindent {\it b) Interacting systems.} It is straightforward to
generalize the proofs to include spin and some class of interactions.
One can accommodate for example a Hubbard term $$
\sum_{x\in \Lambda} h_x= \sum_{x\in L} h_x +\sum_{x\in R} h_x
$$ where $h_x=U(n_{x\uparrow}-{1\over 2})(n_{x\downarrow}-{1\over
2})$, $n_{x\sigma}=c_{x\sigma}^\dagger c_{x\sigma}$,
$\sigma=\uparrow,\downarrow$, $U$ is an arbitrary real number.
Another example is a nearest neighbor repulsive potential 
$$
\sum_{x,y\in\Lambda} h_{xy}= \sum_{x,y\in L} h_{xy}+
\sum_{x,y\in M} h_{xy}
+\sum_{x,y\in R} h_{xy}
\label{Vint}
$$ 
where $h_{xy}=V (n_{x\uparrow}+n_{x\downarrow}-1)
(n_{y\uparrow}+n_{y\downarrow}-1)$ with $V$ a positive number.  Longer
range interactions, and spin dependent forces such as a Heisenberg
antiferromagnetic exchange term can also be included. These cases are
also discussed in \cite{Lie}. Let us describe what happens in the
transformations i)-iii).  In the first step i) one has to replace
\eq{JW} by $$
\dd{x\sigma} = (-1)^{N_L}\cc{x\sigma}\ , \quad
\ddd{x\sigma} = \ccd{x\sigma}(-1)^{N_L}
$$ with $N_L=\sum_{x\in L} (n_{x\uparrow} + n_{x\downarrow})$.  In
step ii) for $x\in R$, $n_{x\sigma}\to 1-n_{x\sigma}$. The Hubbard term
remains unchanged, but the nearest neighbor interaction becomes 
$$
\sum_{x,y\in\Lambda} h_{xy}= \sum_{x,y\in L} h_{xy}-
\sum_{x,y\in M} h_{xy}
+\sum_{x,y\in R} h_{xy} 
$$ 
Thus the interaction between the left
and right parts of the lattice is of the form $\sum_i C_i\otimes C_i$
with the correct sign because $V>0$.  The third step iii) is a gauge
transformation which does not affect the interaction terms.
Summarizing we see that we can bring the Hamiltonians into the form
\eq{TAB}.
Then the proofs of Lemma \ref{lem:reflection} and Theorem
 \ref{thm:main} are unchanged.

We believe that these remarks are useful in other problems.  We
illustrate this by two examples: the $t-V$ model and a generalized
Falicov Kimball model.

\noindent {\it c) Spinless $t-V$ model.} 
This model of spinless electrons has 
Hamiltonian \ref{hams1} with the interaction part equal to
$V\sum_{x,y\in\Lambda} (n_x-1/2)(n_y-1/2)$ where the sum is over
nearest neighbors only and $V$ is positive. The remarks above show
that on a cubic lattice, i.e., $D=3$, 
with periodic boundary conditions in all
directions and a flux configuration through each square plaquette
equal to $\pi$, and $\vert t_{xy}\vert=t$ for all bonds $<xy>$, it can
be brought in a reflection positive form with respect to all
reflection planes. Then it is an exercise to see that the methods of
\cite{DLS} used for the Heisenberg model can be used also in the
present situation to prove that, when $t/V$ is small enough there is
long range order at low temperature $\beta^{-1}$. More precisely if
$<\,\cdot\,>_\Lambda$ is the thermal average with periodic boundary
conditions, one can prove $(-1)^{|x|+|y|}<(n_x-1/2)(n_y-1/2)>_\Lambda
> c > 0$, for all $x$ and $y$ in $\Lambda$, for some strictly positive
constant $c$ independent of $\Lambda$. (We note that in the present
case the uniform density theorem \cite{LLM} applies, so in particular
$<n_x>_\Lambda=1/2$ for all $\beta$, $t$ and $V$). In fact this result
is true for any flux configuration and one can also add a small
chemical potential term (see \cite{Al},\cite{FrFeDa} and
\cite{LeMa} for recent rigorous results). Although our proof does not
work in $D=2$, the result is expected to hold also in two dimensions.

\noindent {\it d)} One can also consider the case of spin 1/2 electrons
with attractive Hubbard interaction and a nearest neighbor repulsion,
i.e., a $t-V-U$ model, and it can be shown that for low enough
temperatures, in three or more dimensions, for $t/V$ small enough, and
$U+4V\leq 0$, the model has checkerboard long range order of the
electron density by following the proof of \cite{DLS} for the
Heisenberg antiferromagnet. Here again, checkerboard long range order
is expected to occur also in $D=2$, but the proof given here does
not directly apply.

\noindent {\it e) Extended Falicov- Kimball model.} The extended
 Falicov-Kimball model we wish to mention has the Hamiltonian
\bea
H(\{w_x\})&= & \sum_{x,y\in\Lambda,\sigma=\uparrow,\downarrow}
t_{xy}c_{x\sigma}^\dagger c_{y\sigma} + U\sum_{x\in \Lambda} h_x
+V\sum_{x,y\in\Lambda} h_{xy}\\
&&+{U^\prime}\sum_{x\in\Lambda}(n_{x\uparrow}+n_{x\downarrow}
-1)(2w_x-1)
\label{faki}
\eea

where $\{w_x\}$ is a configuration of random variables with values $0$
or $1$, describing the position of classical particles (say nuclei or
fermions with a large effective mass, we refer to \cite{KL} for a
 discussion of the
physical interpretations).  The usual Falicov-Kimball model has
$U=V=0$ and only one type of electron (say the spin up electrons). The
energy of a nuclear configuration $\{w_x\}$ is
$\lambda_0(H(\{w_x\}))$, the smallest eigenvalue of \eq{faki} in the
total Fock space of the electrons. A theorem of Kennedy and Lieb
\cite{KL} asserts that for the usual Falicov-Kimball model on a
bipartite lattice $\Lambda=A\cup B$ union of two sublattices $A$ and
$B$, for all $U^\prime$ the minimum of $\lambda_0(H(\{w_x\}))$ is
attained for one of the two configurations $(w_x=0, x\in A, w_x=1,
x\in B)$ or $(w_x=0, x\in B, w_x=1, x\in A)$. This is true
irrespective of the boundary conditions or the flux configuration
(provided it exists). Many more detailed results are known
but it is only this one that we will now generalize.

We take a $D$-dimensional hypercubic lattice with periodic boundary
conditions in all directions and set the flux configuration to be
equal to $\pi$ in all square plaquettes, and also the canonical flux
through the circuits created by the periodic boundary conditions. It
is explained in the next section why this can be done and why it is
the correct choice.  We set $\vert t_{xy}\vert=t$. By performing the
sequence of transformations i)-iii) the Hamiltonian is brought to a
refection positive form. The only term we have not discussed so far is
the last one in \eq{faki} (the one with coupling constant $U^\prime$.)
After the transformations i)-iii) it becomes the following:
\be
{U^\prime}\sum_{x\in L}(n_{x\uparrow}+n_{x\downarrow} -1)(2w_x-1)
-{U^\prime}\sum_{x\in R}(n_{x\uparrow}+n_{x\downarrow} -1)(2w_x-1)
\ee
where $L$ and $R$ refer to the left and right parts of the lattice
with respect to some reflection plane $P$. Let $r(x)$ denote the site
obtained by reflection of $x$ through $P$.  It is convenient to use
the variables $s_x=2w_x - 1$ and write $E_0(\{s_x\}_{x\in L},
\{s_x\}_{x\in R})=\lambda_0(H(\{w_x\}))$.  By applying Lemma
\ref{lem:DLS} one obtains
\be
E_0(\{s_x\}_{x\in L}, \{s_x\}_{x\in R})
\geq \frac{1}{2}\left(E_0(\{s_x\}_{x\in L}, \{s_x^\prime\}_{x\in R})
+E_0(\{s_x^{\prime\prime}\}_{x\in L}, \{s_x\}_{x\in R})
\right)
\ee
where for $x\in R$ $s_x^\prime=-s_{r(x)}$ and for $x\in L$
$s_x^{\prime\prime}=-s_{r(x)}$.  By repeated reflections (across all
reflection planes in all $D$ directions) one concludes that the
minimum energy is attained in the checkerboard configurations of the
variables $s_x$ (or, equivalently, of the $w_x$). Note that this
result holds for all $U,U^\prime$, and for all $V\geq 0$.

\newsection{Examples and Discussions}\label{sec:examples}

In this section we comment and illustrate Theorem \ref{thm:main} by
various examples.  First let us consider several planar graphs.

\noindent {\it 1. Planar graphs}. The most basic case is that of a square
lattice with periodic boundary conditions in one direction and an even
number of sites in that direction, say the horizontal one. Thus we
have a cylinder (which can be embedded in the plane). A basic set of
circuits $\C$ is constituted by the square plaquettes of length $n=4$,
and one big circle along a basis of the cylinder. We emphasize that if
one takes only the square plaquettes then the flux through circuits
that wind around the cylinder is not uniquely determined, and thus the
set of square plaquettes alone is not a basic set of
circuits. Obviously one can find reflection planes that satisfy our
assumptions: these are the vertical planes that cut the cylinder in
two equal halves.  Furthermore in order to have $\vert t_{xy}\vert$
invariant under reflection across these planes we must require that
$\{\vert t_{xy}\vert, <xy>\, \mbox{horizontal}\}$ has period $2$ in
the horizontal direction, and $\{\vert t_{xy}\vert, <xy>\,
\mbox{vertical}\}$ is translation invariant in the horizontal direction.
There is no constraint for $\vert t_{xy}\vert$ along the vertical
direction.  A canonical flux configuration can be described by putting
a flux $\pi$ through each square plaquette and $\pi(N -1)$ through the
basis of length $2N$ of the cylinder. Theorem \ref{thm:main} states
that this flux configuration minimizes the ground state energy.

\begin{figure}[t]
\label{fig:3}
\begin{center}
\epsfysize=8truecm
\centerline{\epsfbox{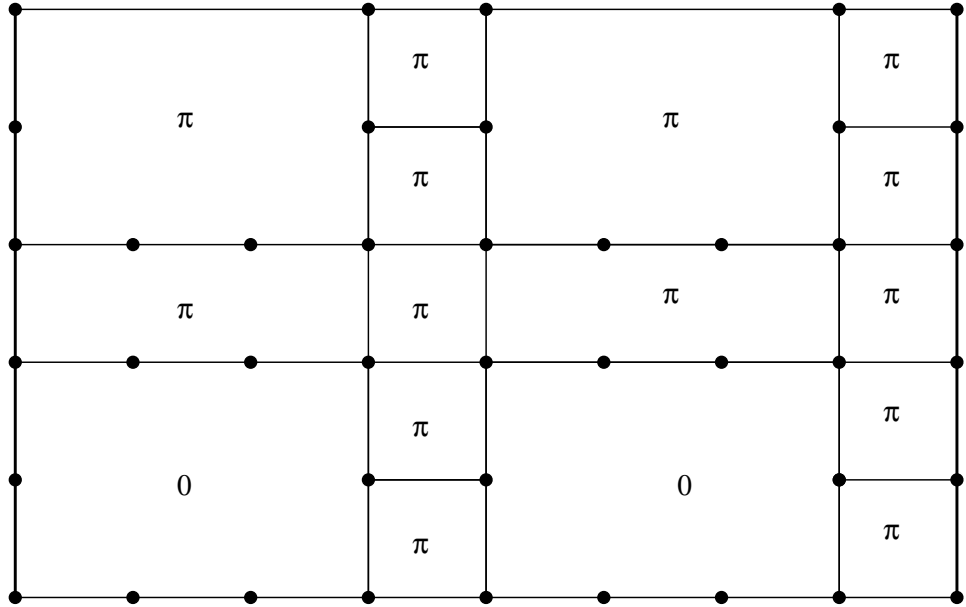}}
\parbox{14truecm}{\caption{\baselineskip=16pt
 A graph with non uniform optimal flux. Periodic boundary
conditions are assumed in the horizontal direction.}}
\end{center}
\end{figure}

As Lieb points out \cite{Lie}, one can obtain the optimal flux on
other graphs simply by "erasing", that is, letting $\vert
t_{xy}\vert\to 0$ on some edges in a way that preserves the
assumptions. For example one can get the hexagonal lattice with periodic
boundary conditions where the flux through each hexagon is $0$ and
$\pi(N -1)$ through the basis. Many planar graphs that cannot be
obtained by this procedure satisfy our assumptions however.  An
example is given in Figure 3.

In all these examples one can also take periodic boundary conditions
in both the vertical and horizontal directions. This wraps the graph
on a torus, thus it is not planar any more. A new circuit has to be
added to $\Cir$, namely a circuit winding around the torus in the
"vertical" direction.  The reason for this will become clear in the
next paragraph.

\noindent {\it 2. Basic sets of circuits}. 
Before discussing more general examples it is useful to indicate a way
of checking that a set $\Cir$ is a basic set of circuits.  We describe
a sufficient condition. We start by representing any oriented circuit
$\gamma$ as a sum over all edges $e_j$ of the graph $$
\gamma=\sum_j\epsilon_j e_j
$$ The edges have a fixed reference orientation and the $\epsilon_j$
are equal to $0$ if $e_j$ does not belong to $\gamma$, $+1$
($resp. -1$) if $e_j$ occurs in $\gamma$ with the same
(respectively. opposite) orientation than $\gamma$.  We require that any
circuit $\gamma$ can be decomposed as
\be
\gamma =\sum_i a_i \gamma_i 
\label{decomposition}
\ee
 with integer $a_i$ and
$\gamma_i\in \Cir$. We call such sets $\Cir$ generating.

If the flux configuration is specified for all $\gamma_i\in \Cir$ then
using \eq{decomposition} we can compute the flux through $\gamma$,
namely $\Phi_\gamma=\sum_i a_i\Phi_i$. Since the flux configuration
through $\gamma_i\in \Cir$ corresponds to a set of phases, different
decompositions of $\gamma$ lead to the same flux.  Once the flux is
determined for all circuits, it follows from Lemma 2.1 in \cite{LL}
that $\Cir$ is a basic set of circuits.

The property that $\Cir$ is generating can be expressed as a simple
topological property of the surface (two-dimensional complex)
consisting of the set of vertices $\Lambda$, the edges in $\Gamma$,
and the set of triangles obtained by triangulation of all the circuits
$\gamma\in \Cir$.  The set $\Cir$ is generating if and only if the
first homology group over the integers of this surface is trivial (see
e.g. \cite{ale}). If one views the complex as a continuous two
dimensional manifold this corresponds to the property that any closed
curve can be contracted to a point.

\noindent {\it 3. Non planar examples.} For non planar graphs it is not
obvious that there exists phases which correspond to the canonical
flux.  Let us consider, e.g., a single $D$-dimensional hypercube. We
show that in general for a given configuration of fluxes through the
two dimensional squares, one cannot find corresponding phases for the
$t_{xy}$.  The number of $k$-dimensional subcubes is equal to $2^{D-k}
D!/ (D-k)! k!$.  Indeed, a $k$ dimensional subcube is determined by
the set of points $(x_1,...,x_D)$ with $0\leq x_{i_1}\leq 1$, ...,
$0\leq x_{i_k}\leq 1$, and $x_j= 0$ or $1$ for $j\neq i_1...i_k$. So
we have $D!/(D-k)! k!$ choices for $i_1...i_k$ and $2^{D-k}$ choices
for the $x_j$'s. Thus the number of flux variables through squares is
$2^{D-3}D(D-1)$, and the number of phases on the edges is
$2^{D-1}D$. In general one will have to solve a system of equations
which is overdetermined if $2^{D-1}D < 2^{D-3}D(D-1)$, i.e. $D>5$.
However for the canonical flux configuration there always exist a
solution of this system of equations. In fact our proof of the flux
phase conjecture constructs such a solution, for any graph satisfying
the assumptions of Theorem \ref{thm:main}.

In particular the hypercubic lattice falls into our class of
graphs. In order to satisfy the assumptions we have to take periodic
boundary conditions in $D-1$ or $D$ directions. A generating set of
circuits is constituted by all the square plaquettes and $D-1$
circuits, that are the $D-1$ coordinate axis in the periodic
directions. The canonical flux configuration is unique and equals
$\pi$ for each plaquette and $\pi (N_i-1)$ through the $D-1$ circuits
in the periodic directions of lengths $2N_i$, $i=1,...,D-1$.  One can
of course imagine many non-planar graphs satisfying the assumptions
A1-A2.

\section*{Acknowledgements}

The authors wish to thank C. Baesens, and R. MacKay for organising
the Electron - Phonon Workshop in Warwick (September 1994), where this 
work was initiated.
B.N. would like to thank the Institut de Physique Theorique at EPF,
Lausanne, where part of this paper was written, for kind hospitality,
and both authors thank the Centre de Physique Th\'eorique,
 Luminy, where this work was
completed, for a most enjoyable stay. We also thank Almut
Burchard, Pirmin Lemberger, Elliott Lieb, Alain Messager, 
Salvador Miracle-Sol\'e, and
Jean Ruiz, for interesting discussions.  The work of B.N. is partially
supported by the U.S.  National Science Foundation under Grant
No. PHY90-19433 A04.

\def\thebibliography#1{\section*{References}\list
  {\arabic{enumi}.}{\settowidth\labelwidth{[#1]}
    \leftmargin\labelwidth \advance\leftmargin\labelsep
    \usecounter{enumi}} \def\newblock{\hskip .11em plus .33em minus
    -.07em} \sloppy \sfcode`\.=1000\relax}

\end{document}
\bye